\documentclass{article}
\usepackage{amssymb}

\input{tcilatex}

\begin{document}

\title{Discrete, q-difference deformations of associative algebras and
integrable systems}
\author{ B.G. Konopelchenko \\
\\
Dipartimento di Fisica, Universita del Salento \\
and INFN, Sezione di Lecce, 73100 Lecce, Italy}
\maketitle

\begin{abstract}
\bigskip

Discrete and q-difference deformations of the structure constants for a
class of associative noncommutative algebras are studied. It is shown that
these deformations are governed by a central system of discrete or
q-difference equations which in particular cases represent discrete and
q-difference versions of the oriented associativity equation. It is
demonstrated also that the celebrated Hirota-Miwa bilinear equation for the
AKP and BKP hierarchies describes discrete deformations of certain
finite-dimensional algebras.
\end{abstract}

\section{Introduction}

One of the approaches within the deformation theory for
associative algebras proposed by Gerstenhaber in his seminal
papers [1,2] consists in the treatment of '' the set of structure
constants of an algebra in a given basis as parameter space for
the deformation theory''. A remarkable class of such deformations
was discovered by Witten [3], Dijkgraaf-Verlide-Verlinde [4] and
beautifully formalized by Dubrovin [5,6] in terms of Frobenius
manifolds and subsequently by Hertling and Manin [7,8] in terms of
F-manifolds (see also [9,10]).

A different method to describe classes of deformations for associative
algebras, namely, coisotropic and quantum deformations has been proposed
recently in [11-13]. For the quantum deformations [13] this approach
consists 1) in putting the correspondence between the table of
multiplication for an associative algebra in the basis $\mathbf{P}_{0},%
\mathbf{P}_{1},...,\mathbf{P}_{N-1},$ i.e.

\begin{equation}
\mathbf{P}_{j}\mathbf{P}_{k}=C_{jk}^{l}\mathbf{P}_{l}, \quad
j,k=0,1,...N-1
\end{equation}
and the set of operators

\begin{equation}
f_{jk}=-p_{j}p_{k}+C_{jk}^{l}(x^{0},x^{1},...,x^{N-1})p_{l},\quad
j,k=0,1,...,N-1
\end{equation}%
where $x^{0},x^{1},...,x^{N-1}$ stand for the deformation parameters and
summation over repeated index ( from 0 to N-1 ) is assumed, 2) the
requirement that the operators $p_{0},p_{1},...,p_{N-1}$and $%
x^{0},x^{1},...,x^{N-1}$ are elements of the Heisenberg algebra, i.e.

\begin{equation}
\lbrack p_{j},p_{k}]=0, \quad [x^{j},x^{k}]=0, \quad
[p_{j},x^{k}]=\hbar \delta _{j}^{k}, \quad j,k=0,1,...,N-1
\end{equation}
where $\hbar $ is a constant and $\delta _{j}^{k}$ is the Kronecker symbol
and 3) the requirement that the functions $C_{jk}^{l}(x)$ are such that the
set of equations

\begin{equation}
f_{jk}\mid \Psi \rangle =0, \quad j,k=0,1,...,N-1
\end{equation}
has a nontrivial common solution ( Dirac's prescription) where $\mid \Psi
\rangle $ are elements of a certain linear space.

The requirement (4) gives rise to the set of equations ( quantum central
system (QCS)) [13]

\begin{equation}
\hbar \frac{\partial C_{jk}^{n}}{\partial x^{l}}-\hbar \frac{\partial
C_{kl}^{n}}{\partial x^{j}}+C_{jk}^{m}C_{lm}^{n}-C_{kl}^{m}C_{jm}^{n}=0,%
\quad j,k,l,n=0,1,...,N-1
\end{equation}%
which governs quantum deformations of the structure constants $%
C_{jk}^{l}(x^{0},...,x^{N-1})$ . The QCS (5) has a simple geometrical
meaning of vanishing Riemann curvature tensor and \ contains the oriented
associativity equation, WDVV equation, Boussinesq equation, Gelfand-Dikii
and Kadomtsev-Petviashvili (KP) hierarchies as the particular cases [13].

\ In the present paper we will define and study discrete and q-difference
deformations of associative algebras. The basic steps of the construction
are quite similar to those for quantum deformations. First, we ''identify''
the elements $\mathbf{P}_{0},\mathbf{P}_{1},...,\mathbf{P}_{N-1}$ of a basis
and deformation parameters $x^{0},x^{1},...,x^{N-1}$ with the elements of
the algebra of shifts or q-shifts. Then, the requirement of existence of
nontrivail common solutions of equations (4) provides us with the central
system (DCS) which governs discrete or q-difference deformations of the
structure constants $C_{jk}^{l}(x).$ This DCS is the discrete or
q-difference version of the QCS (5). As the particular cases it contains the
discrete and q-difference versions of the oriented associativity equations
and other integrable equations. The construction provides us with the
discrete version of the curvature tensor connected in a simple way with the
associator of the algebra.

\ We demonstrate also that the discrete deformations of an algebra for which
the multiplication of only distinct elements are admited are described, in
particular, by the discrete Darboux system and by the famous bilinear
Hirota-Miwa equations for the AKP and BKP hierarchies.

\ The paper is organized is follows. In section 2 discrete and q-difference
deformations of associative algebras are defined and the corresponding DCSs
are derived. Discrete associator, discrete version of the curvature tensor
and their interrelation are discussed in section 3. Reduction of the DCS to
the discrete versions of the oriented associativity equation are studied in
section 4. Discrete deformations governed by the Hirota-Miwa bilinear
equations are considered in section 5.

\section{Discrete and q-difference deformations}

Thus, we consider a finite-dimensional associative noncommutative algebra
\textit{A }with (or without) unite element $\mathbf{P}_{0}$ . We will
restrict ourselfs to a class of algebras which possess a basis composed by
pairwise commuting elements. Denoting elements of such a basis as $\mathbf{P}%
_{0},\mathbf{P}_{1},...,\mathbf{P}_{N-1}$ , one has the corresponding
multiplication table

\begin{equation}
\mathbf{P}_{j}\mathbf{P}_{k}=C_{jk}^{l}\mathbf{P}_{l}, \quad
j,k=0,1,...,N-1
\end{equation}
The commutativity of for the basis implies that $C_{jk}^{l}=C_{kj}^{l}$.

In order to define deformations $C_{jk}^{l}(x^{0},x^{1},...,x^{N-1})$ of the
structure constants we first identify the elements of the basis $P_{j}$ and
deformation parameters $x^{j}$ with the elements of the algebra defined by
the commutation relations

\begin{equation}
\lbrack p_{j},p_{k}]=0,\quad \lbrack x^{j},x^{k}]=0,\quad \lbrack
p_{j},x^{k}]=\delta _{j}^{k}(\hat{I}+p_{j}),\qquad j,k=0,1,...,N-1
\end{equation}%
where $\hat{I}$ denotes the identity operator. It is easy to notice that the
algebra of shifts, i.e. $\Delta _{j}=T_{j}-1,T_{j}(x^{k})=x^{k}+\delta
_{j}^{k},j,k=0,1,...,N-1$ with the rule $\Delta _{j}(f\cdot g)=\Delta
_{j}f\cdot g+T_{j}f\cdot \Delta _{j}g$ is \ a realization of this algebra: $%
p_{j}=\Delta _{j}.$

Next steps are to introduce the operators

\begin{equation}
f_{jk}\doteqdot -p_{j}p_{k}+C_{jk}^{l}(x)p_{l}, \quad
j,k=0,1,...,N-1
\end{equation}
and to require that the functions $C_{jk}^{l}(x)$ are such that these
operators have common nontrivial kernel or, equivalently, that the equations

\begin{equation}
f_{jk}\mid \Psi \rangle =0, \quad j,k=0,1,...,N-1
\end{equation}
are compatible. Here $\mid \Psi \rangle $ are elements of a linear space
where the action of the operators $p_{j}$ and $x^{j}$ is defined.

\textbf{Definition.} The structure constants $C_{jk}^{l}(x)$ of the
associative algebra \textit{A }are said to define deformations generated by
the algebra (7) if the operators $f_{jk}$ given by (8) have a common
nontrivial kernel.

This definition can be converted into the set of equations for the structure
constants. The basic tools are given by the following two identities. The
first is

\begin{equation}
\lbrack p_{j},\varphi (x)]=\Delta _{j}\varphi (x)\cdot
(\hat{I}+p_{j}), \quad j=1,2,...,N-1
\end{equation}
where $\varphi (x)$ is an arbitrary function, $\Delta _{j}\varphi
(x^{0},x^{1},...,x^{N-1})=(T_{j}-1)\varphi (x^{0},x^{1},...,x^{N-1})$ and $%
T_{j}\varphi (x^{0},...,x^{j},...,x^{N-1})=\varphi
(x^{0},...,x^{j}+1,....,x^{N-1}).$ The second is

\begin{eqnarray}
\left( p_{j}p_{k}\right) p_{l}-p_{j}\left( p_{k}p_{l}\right)
=-p_{l}f_{jk}+p_{j}f_{kl}-T_{l}C_{jk}^{m}\cdot
f_{lm}+T_{j}C_{kl}^{m}\cdot f_{jm}+ \quad &
\nonumber
\\
 + (\Delta _{l}C_{jk}^{n}-\Delta
_{j}C_{kl}^{n}+T_{l}C_{jk}^{m}\cdot
C_{lm}^{n}-T_{j}C_{kl}^{m}\cdot C_{jm}^{n})p_{n},\quad
j,k,l=1,2,...,N-1
\end{eqnarray}
The identity (11) implies that

\begin{eqnarray}
\left\{ \left( p_{j}p_{k}\right) p_{l}-p_{j}\left(
p_{k}p_{l}\right) \right\} \mid \Psi \rangle
=A_{klj}^{n}(x)p_{n}\mid \Psi \rangle =\quad &
\nonumber
\\
 = \left( \Delta _{l}C_{jk}^{n}-\Delta
_{j}C_{kl}^{n}+T_{l}C_{jk}^{m}\cdot
C_{lm}^{n}-T_{j}C_{kl}^{m}\cdot C_{jm}^{n}\right) p_{n}\mid \Psi
\rangle
\end{eqnarray}
where $\mid \Psi \rangle \subset $ linear subspace $\mathit{H}_{\Gamma }$
defined by equations (9) and $A_{klj}^{n}$ is the associator for the algebra%
\textit{\ A. \ }Weak associativity, i.e. the requirement that l.h.s. of (12)
vanishes, means that the \ r.h.s. of (16) should vanishes too. \ This is
valid for all values of the deformation parameters if

\begin{equation}
A_{klj}^{n}=\Delta _{l}C_{jk}^{n}-\Delta
_{j}C_{kl}^{n}+T_{l}C_{jk}^{m}\cdot
C_{lm}^{n}-T_{j}C_{kl}^{m}\cdot C_{jm}^{n}=0, \quad
j,k,l,n=0,1,...,N-1
\end{equation}
If the subspace $\mathit{H}_{\Gamma }$ does not contain elements linear in $%
p_{j}\mid \Psi \rangle $ , then equation (13) represents also the necessary
condition.

Thus, we have

\textbf{Proposition. }\ The structure constants $C_{jk}^{l}(x)$ define
deformations driven by the algebra (7) if they obey equation (13).

We will refer to the system (13) as the discrete central system (DCS) . We
note \ that though the DCS (13) seems directly connected with the algebra of
shifts, it defines deformations generated by the abstract algebra (7).

In a similar manner one defines q-difference deformations of associative
algebras. Considering the same class of associative noncommutative algebras
and \ following the same scheme , one , instead \ \ of the algebra (7),
takes the algebra of q-shifts, i.e.

\begin{equation}
\lbrack p_{j},p_{k}]=0, \quad [x^{j},x^{k}]=0, \quad
[p_{j},x^{k}]=\delta _{j}^{k}(\hat{I}+qx^{j}p_{j}), \quad
j,k=0,1,...,N-1
\end{equation}
where q is an arbitrary number. A realization of this algebra is given by

\begin{equation}
p_{j}=\Delta _{qj}=\frac{1}{qx^{j}}\left( T_{qj}-1\right)
\end{equation}
where

\[
T_{qj}(x^{k})=x^{k}+q\delta _{j}^{k}x^{k}, \quad j,k=0,1,...,N-1
\]
and

\[
T_{qj}\varphi (x^{0},...,x^{N-1})=\varphi
(x^{0},...,(1+q)x^{j},...,x^{N-1}).
\]
Since

\[
\Delta _{qj}(f\cdot g)=\Delta _{qj}f\cdot g+T_{qj}f\cdot \Delta _{qj}g
\]
the central system which govers the deformations driven by the algebra (14)
is quite similar to the DCS (13). It is

\begin{equation}
\Delta _{ql}C_{jk}^{n}-\Delta _{qj}C_{kl}^{n}+T_{ql}C_{jk}^{m}\cdot
C_{lm}^{n}-T_{qj}C_{kl}^{m}\cdot C_{jm}^{n}=0.
\end{equation}
In spite of this similarity solutions of the central systems (13) and (16)
are rather different. So, algebras (7) and (14) generate quite different
deformations of the structure constants for the same algebra.

We will refer to the algebra of the type (7) and (14) which generates
deformations of associative algebra within our scheme as the Deformation
Driving Algebra (DDA) to avoid possible confusion with the other already
existing abbreviations like DGA( Deformation Generating Algebra) and so on
(see e.g. [14]).

Similar to the other cases one can presents the DCS (13) in a compact matrix
form. Using the standard matrices $C_{j}$ ,$A_{lj}$ defined by $\left(
C_{j}\right) _{k}^{l}=C_{jk}^{l}$ and $\left( A_{lj}\right)
_{k}^{n}=A_{klj}^{n}$ , one rewrites the DCS (13) and (16) as

\begin{equation}
A_{lj}^{d}\doteqdot \Delta _{l}C_{j}-\Delta
_{j}C_{l}+C_{l}T_{l}C_{j}-C_{j}T_{j}C_{l}=0
\end{equation}
and

\begin{equation}
A_{lj}^{q}\doteqdot \Delta _{ql}C_{j}-\Delta
_{qj}C_{l}+C_{l}T_{ql}C_{j}-C_{j}T_{qj}C_{l}=0
\end{equation}
or

\begin{equation}
A_{lj}^{d}=(1+C_{l})T_{l}(1+C_{j})-(1+C_{j})T_{j}(1+C_{l})=0
\end{equation}
and

\[
A_{lj}^{q}=(1+C_{l})T_{ql}(1+C_{j})-(1+C_{j})T_{qj}(1+C_{l})=0.
\]

At last, we would like to notice that the DCS (13) which is the consequence
of the weak associativity condition

\begin{equation}
\left\{ \left( p_{j}p_{k}\right) p_{l}-p_{j}\left( p_{k}p_{l}\right)
\right\} \mid \Psi \rangle =0
\end{equation}
at the realization $p_{j}=\Delta _{j}$ eventually coincides with the
compatibility condition for the linear problems $f_{jk}\mid \Psi \rangle =0$%
, i.e.

\[
\left\{ \Delta _{j}\Delta _{k}-C_{jk}^{l}(x)\Delta _{l}\right\} \mid \Psi
\rangle =0.
\]

\section{Associator, discrete curvature and linear problems.}

In the continuous limit $\Delta _{j}\rightarrow \varepsilon \hbar \frac{%
\partial }{\partial x^{j}},C_{j}\rightarrow \varepsilon C_{j},\varepsilon
\rightarrow 0$ all the above equations are reduced to those for quantum
deformations [13]. In particular, the DCS (13) ( or (17)) is converted into
the QCS (5) which has the geometrical meaning of vanishing \ Riemann
curvature tensor $R_{klj}^{n}$ given by the l.h.s. of equation (5) with the
structure constants $C_{jk}^{l}$ identified with the Christoffel symbols. We
recall that in the continuous case the matrix \ $R_{jk}^{class}$ with the
matrix elements $\left( R_{jk}^{class}\right) _{l}^{n}=R_{ljk}^{n}$ is the
commutator

\begin{equation}
\ R_{jk}^{class}=[\nabla _{j},\nabla _{k}], \quad j,k=0,1,...,N-1
\end{equation}
where $\nabla _{j}=\hbar \frac{\partial }{\partial x^{j}}+C_{j}$ and the
equation \ $R_{jk}^{class}=0$ is equivalent to the compatibility condition
for the linear problems

\begin{equation}
\nabla _{j}\Psi =\left( \hbar \frac{\partial }{\partial
x^{j}}+C_{j}\right) \Psi =0,\quad j=0,1,...,N-1.
\end{equation}%
In the continuous \ case the corresponding version of \ the relation (12)
implies that the associator $A_{jk}^{class}$ coincides with the Riemann
curvature matrix $R_{jk}^{class}$.

The situation is quite different in the discrete case. The discrete
associator $A_{lj}^{d}$ is given by the formula (17) or (19). \ In order to
introduce the discrete analog of the curvature tensor we observe that
equations (19) are equivalent to the compatibility condition for the linear
system

\begin{eqnarray}
L_{j}\mid \Psi \rangle =\left( T_{j}-(1+C_{j})^{-1}\right) \mid
\Psi \rangle =\left( \Delta _{j}-(1+C_{j})^{-1}C_{j}\right) \mid
\Psi \rangle =0, \quad &
\nonumber
\\
j=0,1,...,N-1.
\end{eqnarray}
Indeed, in virtue of the relation

\begin{eqnarray}
\lbrack L_{j},L_{k}]\mid \Psi \rangle =\left\{
T_{k}(1+C_{j})^{-1}\cdot (1+C_{k})^{-1}-T_{j}(1+C_{k})^{-1}\cdot
(1+C_{j})^{-1}\right\} \mid \Psi \rangle,
\nonumber
\\
j,k=0,1,...,N-1
\end{eqnarray}
equations (23) are compatible if

\begin{equation}
T_{k}(1+C_{j})^{-1}\cdot (1+C_{k})^{-1}-T_{j}(1+C_{k})^{-1}\cdot
(1+C_{j})^{-1}=0, \quad j,k=0,1,...,N-1.
\end{equation}
These equations are obviously equivalent to equations (19) provided all
matrices $1+C_{j}$ are nondegenerate. Here and in the rest of this section
there is no summation over repeated indices. The relation (24) in analogy
with the continuous case suggests to treat the expression in the bracket of
the r.h.s., i.e.

\begin{eqnarray}
R_{jk}^{d}=T_{k}(1+C_{j})^{-1}\cdot
(1+C_{k})^{-1}-T_{j}(1+C_{k})^{-1}\cdot (1+C_{j})^{-1}=\quad &
\nonumber
 \\
  = \Delta _{k}(1+C_{j})^{-1}\cdot (1+C_{k})^{-1}-\Delta
_{j}(1+C_{k})^{-1}\cdot
(1+C_{j})^{-1}+[(1+C_{j})^{-1},(1+C_{k})^{-1}]
\end{eqnarray}
as the curvature ''tensor''.The relation (24) provided us with the weak
definition of the curvature ''tensor''

\begin{equation}
R_{jk}^{d}\mid \Psi \rangle =[L_{j},L_{k}]\mid \Psi \rangle ,
\quad j,k=0,1,...,N-1.
\end{equation}
Using the explicit form of $L_{j},$ one derives the following operator
expression for the discrete curvature ''tensor''

\begin{equation}
R_{jk}^{d}=[L_{j},L_{k}]+\Delta _{j}(1+C_{k})^{-1}\cdot
L_{j}-\Delta _{k}(1+C_{j})^{-1}\cdot L_{k},\quad j,k=0,1,...,N-1.
\end{equation}
Comparing the formulae (19) and (26) , one also concludes that

\begin{equation}
A_{jk}^{d}=(1+C_{j})T_{j}(1+C_{k})\cdot R_{jk}^{d}\cdot
(1+C_{k})T_{k}(1+C_{j}).
\end{equation}
In the continuous limit $T_{j}=1+\varepsilon \frac{\partial }{\partial x^{j}}%
,C_{j}\rightarrow \varepsilon C_{j},(1+C_{j})^{-1}\rightarrow 1-\varepsilon
C_{j},L_{j}\rightarrow \varepsilon (\frac{\partial }{\partial x^{j}}+C_{j})$
and

\begin{equation}
R_{jk}^{d} \rightarrow \varepsilon ^{2}[\nabla _{j},\nabla _{k}]+\varepsilon
^{3}...=\varepsilon ^{2}R_{jk}^{class}+\varepsilon ^{3}...,
\end{equation}
\begin{equation}
A_{jk}^{d} \rightarrow \varepsilon ^{2}A_{jk}^{class}+...,
\end{equation}
and

\[
A_{jk}^{class}=R_{jk}^{class}.
\]

The above formulae also indicate that in the situation in which one ignores
the relation with associative algebras , it is natural to define a discrete
curvature '' tensor'' associated the discrete ''connection'' \ $L_{j}=\Delta
_{j}+B_{j}$ \ as follows

\begin{eqnarray}
R_{jk}^{d}=[L_{j},L_{k}]-\Delta _{j}B_{k}\cdot L_{j}+\Delta
_{k}B_{j}\cdot L_{k}= &
\nonumber
\\
= \Delta _{j}B_{k}\cdot (1-B_{k})-\Delta _{k}B_{j}\cdot
(1-B_{j})+[B_{j},B_{k}],  & \nonumber
\\ j,k=0,1,...,N-1.
\end{eqnarray}
The connection with the original matrices $C_{j}$ is given by $%
(1+C_{j})(1-B_{j})=1.$

Finally, we present the linear problems with the spectral parameter $\lambda
$ for the DCS (19). It is

\begin{equation}
L_{j}(\lambda )\mid \Psi \rangle =\left( T_{j}-\lambda (1+C_{j})^{-1}\right)
\mid \Psi \rangle =0.
\end{equation}

\section{Discrete oriented associativity equation}

\ For general discrete or q-difference deformations , similar to the quantum
deformations [13], the global associativity condition $[C_{j},C_{k}]=0$ is
not preserved for all values of the deformation parameters. Deformations of
associative algebras for which the associativity condition is globally valid
( isoassociative deformations) form an important class of all possible
deformations [5-13]. Within the theories of Frobenius and F-manifolds \
[5-10] and also for the coisotropic and quantum deformations [11-13] such
deformations are characterized by the existence of a set of functions $\Phi
^{l},l=0,1,...,N-1$ such that

\begin{equation}
C_{jk}^{l}=\frac{\partial ^{2}\Phi ^{l}}{\partial x^{j}\partial
x^{k}}, \quad j,k,l=0,1,...,N-1.
\end{equation}

These functions obey the oriented associativity equation [5,15]

\begin{equation}
\frac{\partial ^{2}\Phi ^{n}}{\partial x^{l}\partial x^{m}}\frac{\partial
^{2}\Phi ^{m}}{\partial x^{j}\partial x^{k}}-\frac{\partial ^{2}\Phi ^{n}}{%
\partial x^{j}\partial x^{m}}\frac{\partial ^{2}\Phi ^{m}}{\partial
x^{l}\partial x^{k}}=0, \quad j,k,l,n=0,1,...,N-1.
\end{equation}

Here we will present discrete versions of this equation. So, we consider the
isoassociative \ deformations for which

\begin{equation}
\lbrack C_{j}(x),C_{k}(x)]=0, \quad j,k=0,1,...,N-1.
\end{equation}
For such deformations the DCS (17) takes the form

\begin{equation}
\Delta _{l}C_{j}-\Delta _{j}C_{l}+C_{l}\Delta _{l}C_{j}-C_{j}\Delta
_{j}C_{l}=0
\end{equation}
or

\begin{equation}
(1+C_{l})\Delta _{l}(1+C_{j})-(1+C_{j})\Delta _{j}(1+C_{l})=0.
\end{equation}
There is no summation over repeated indices in the formulae (37), (38) and
also in the formula (43). General solution of these equations is

\begin{equation}
C_{j}=g^{-1}\Delta _{j}g
\end{equation}
where g(x) is a matrix-valued function. Since $C_{jk}^{l}=C_{kj}^{l}$ one
has
\begin{equation}
\Delta _{j}g_{k}^{n}=\Delta _{k}g_{j}^{n}, \qquad j,k,n=0,1,...,N-1
\end{equation}
and hence

\begin{equation}
g_{k}^{n}=g_{0k}^{n}+\alpha \Delta _{k}\Phi ^{n}, \qquad k,n=0,1,...,N-1
\end{equation}
where $g_{0k}^{n}$ and $\alpha $ are arbitrary constants and $\Phi ^{n}$ are
functions. Substitution of (39) and (41) into (36) gives

\begin{equation}
\Delta _{l}\Delta _{t}\Phi ^{n}\cdot (g^{-1})_{m}^{t}\Delta
_{j}\Delta _{k}\Phi ^{m}-\Delta _{l}\Delta _{t}\Phi ^{n}\cdot
(g^{-1})_{m}^{t}\Delta _{j}\Delta _{k}\Phi ^{m}=0, \quad
j,k,l,n=0,1,...,N-1.
\end{equation}
Since in the continuous limit $\Delta _{j}\rightarrow \varepsilon \frac{%
\partial }{\partial x^{j}},g_{0k}^{n}=\delta _{k}^{n},\alpha =0,\varepsilon
\rightarrow 0$ the system (42) is reduced to (35), it represents a discrete
isoassociative version of the oriented associativity equation.

Different discrete version of equation (35) arises if one relaxes the
condition (36) and requires that the following quasi-associativity condition

\begin{equation}
C_{l}T_{l}C_{j}=C_{j}T_{j}C_{l}, \quad j,l=0,1,...,N-1
\end{equation}
is valid for all values of deformation parameters . In this case the DCS
(17) is reduced to the system

\[
\Delta _{l}C_{j}-\Delta _{j}C_{l}=0, \quad j,l=0,1,...,N-1
\]
which implies the existence of the matrix-valued function $\Phi $ such that

\[
C_{j}=\Delta _{j}\Phi , \quad j=0,1,...,N-1.
\]
Since $C_{jk}^{l}=\Delta _{j}\Phi _{k}^{l}=C_{kj}^{l}$ one has

\[
\Phi _{k}^{l}=\Delta _{k}\Phi ^{l}, \quad l,k=0,...,N-1
\]
where $\Phi ^{l},l=0,1,...,N-1$ are functions. So

\[
C_{jk}^{l}=\Delta _{j}\Delta _{k}\Phi ^{l}.
\]

Finally, the quasi-associativity condition (43) takes the form

\begin{equation}
\Delta _{j}\Delta _{k}T_{l}\Phi ^{m}\cdot \Delta _{l}\Delta
_{m}\Phi ^{n}-\Delta _{l}\Delta _{k}T_{j}\Phi ^{m}\cdot \Delta
_{j}\Delta _{m}\Phi ^{n}=0, \quad j,k,l,n=0,1,...,N-1.
\end{equation}
which is a discrete version of the oriented associativity equation (35). Any
solution of the systems (42) and (44) defines discrete deformation of the
structure constants $C_{jk}^{l}.$

In a similar manner one derives \ q-difference versions of the oriented
associativity equation.

\section{Hirota-Miwa bilinear equations, discrete Darboux system and
discrete deformations}

Here we will study discrete deformations of algebras for which products of
only distinct elements of the basis $\mathbf{P}_{0},\mathbf{P}_{1},...,%
\mathbf{P}_{N-1}$ are defined and the table of multiplication is of the form

\begin{equation}
\mathbf{P}_{j}\mathbf{P}_{k}=C_{jk}^{k}\mathbf{P}_{k}+C_{jk}^{j}\mathbf{P}%
_{j}+C_{jk}^{0}\mathbf{P}_{0,} \quad j\neq k,j,k=0,1,...,N-1.
\end{equation}
Deformation driving algebra is given by the commutation relations

\begin{equation}
\lbrack p_{j},p_{k}]=0, \quad [x^{j},x^{k}]=0, \quad
[p_{j},x^{k}]=\delta _{j}^{k}p_{j}, \quad j,k=0,1,...,N-1.
\end{equation}
Algebra of shifts $p_{j}=T_{j}$ gives a realization of this abstract
algebra. For any three distinct indices j,k,l one has a closed subtable
(45). Denoting these three indices as 1,2,3, we present a corresponding
subtable of multiplication as

\begin{equation}
\mathbf{P}_{1}\mathbf{P}_{2} =A\mathbf{P}_{1}+B\mathbf{P}_{2}+L\mathbf{P}%
_{0},
\end{equation}
\begin{equation}
\mathbf{P}_{1}\mathbf{P}_{3} =C\mathbf{P}_{1}+D\mathbf{P}_{3}+M\mathbf{P}%
_{0},
\end{equation}
\begin{equation}
\mathbf{P}_{2}\mathbf{P}_{3} =E\mathbf{P}_{2}+G\mathbf{P}_{3}+N\mathbf{P}%
_{0}.
\end{equation}

Discrete central system for the structure constants A,B,...,N is given by
the system of equations

\begin{eqnarray}
\frac{A_{3}}{A} =\frac{C_{2}}{C}, \qquad \frac{B_{3}}{B}=\frac{E_{1}}{E},
\qquad \frac{D_{2}}{D}=\frac{G_{1}}{G}, \\
(A_{3}-E_{1})L+B_{3}N-G_{1}M =0, \qquad (A_{3}-G_{1})D-E_{1}A-N_{1}=0, \\
(A_{3}-G_{1})D+B_{3}G+L_{3} =0, \qquad (C_{2}-E_{1})L+D_{2}N-G_{1}M=0, \\
(C_{2}-E_{1})B+D_{2}E+M_{2} =0, \qquad C_{2}L-A_{3}M+(D_{2}-B_{3})N=0
\end{eqnarray}
where we denote $A_{j}=T_{j}A,B_{j}=T_{j}B$ and so on. \ Equations
(50) imply that there exist three functions U,V,W such that

\begin{equation}
A =\frac{U_{2}}{U},B=\frac{V_{1}}{V},C=\frac{U_{3}}{U}, \\
D =\frac{W_{1}}{W},E=\frac{V_{3}}{V},G=\frac{W_{2}}{W}.
\end{equation}

We will consider here three different reductions of this general system. The
first reduction is associated with the constraints

\begin{equation}
A+B+L =1,
\end{equation}
\begin{equation}
C+D+M =1,
\end{equation}
\begin{equation}
E+G+N =1.
\end{equation}
In the terms of the functions $H^{1},H^{2},H^{3}$ defined by

\begin{equation}
U=H_{1}^{1},V=H_{2}^{2},W=H_{3}^{3}
\end{equation}
the DCS (50-53) under the constraints (55-57) takes the form

\begin{equation}
H_{lk}^{j}-\frac{H_{kl}^{k}}{H_{k}^{k}}H_{k}^{j}-\frac{H_{kl}^{l}}{H_{l}^{l}}%
H_{l}^{j}+\frac{H_{lk}^{k}}{H_{k}^{k}}H^{j}+\frac{H_{lk}^{l}}{H_{l}^{l}}%
H^{j}-H^{j}=0
\end{equation}
or equivalently

\begin{equation}
\Delta _{l}\Delta _{k}H^{j}-\frac{\Delta _{l}H_{k}^{k}}{H_{k}^{k}}\cdot
\Delta _{k}H^{j}-\frac{\Delta _{k}H_{l}^{l}}{H_{l}^{l}}\cdot \Delta
_{l}H^{j}=0
\end{equation}
where indices j,k,l=1,2,3 are all distinct. It is the well-known discrete
Darboux system which was first derived in [16] and then found various
applications in the discrete geometry ( see e.g. [17,18]).

For the general n-dimensional case (45) with the constraints

\begin{equation}
C_{jk}^{k}+C_{jk}^{j}+C_{jk}^{0}=1,\quad j\neq k,
\end{equation}
one has

\begin{equation}
C_{jk}^{k}=\frac{H_{kj}^{k}}{H_{k}^{k}},\quad C_{jk}^{j}=\frac{H_{kj}^{j}}{%
H_{j}^{j}}, \quad C_{jk}^{0}=1-\frac{H_{kj}^{k}}{H_{k}^{k}}-\frac{H_{kj}^{j}%
}{H_{j}^{j}}
\end{equation}
and the DCS is given by equations (59) or (60).

So, the discrete Darboux system describes discrete deformations of the
structure constants for the algebras of the above type. Interrelation
between such algebras and geometrical constructions for the quadrilateral
lattices will be discussed elsewhere.

\ Second reduction is given by the constraints

\begin{equation}
L =M=N=0,
\end{equation}
\begin{equation}
A+B =0,\quad C+D=0,\quad E+G=0.
\end{equation}
Under these constraints the DCS (50-53) becomes

\begin{equation}
\frac{A_{3}}{A} =\frac{C_{2}}{C}=\frac{E_{1}}{E},
\end{equation}
\begin{equation}
A_{3}C+E_{1}C-E_{1}A =0.
\end{equation}

Equations (65) imply that there exists a function $\tau $ such
that

\begin{equation}
A=-\frac{\tau _{1}\tau _{2}}{\tau \tau _{12}}, \quad C=-\frac{\tau
_{1}\tau _{3}}{\tau \tau _{13}}, \quad E=-\frac{\tau _{2}\tau
_{3}}{\tau \tau _{23}}
\end{equation}
where $\tau _{j}=T_{j}\tau $ etc. \ In terms of the function $\tau
$ equation (66) takes the form

\begin{equation}
\tau _{1}\tau _{23}-\tau _{2}\tau _{13}+\tau _{3}\tau _{12}=0.
\end{equation}

It is the famous Hirota discrete bilinear equation for the
Kadomtsev-Petviashvili (AKP) hierarchy [19]. Thus, the Hirota
bilinear equation governs the discrete deformations of the
structure constants from the table of multiplication (47-49) under
the constraints (63-64).

\ It is worth to note that these deformations are isoassociative
one. \ Indeed, the only associativity condition for the
''algebra'' (47-49) with the constraints (63-64) is given by the
relation

\begin{equation}
AC+EC-AE=0.
\end{equation}
The Hirota equation (68) is exactly the associativity condition
(69) with the structure constants A,C,E given by the formulae
(67).

\ \ The third reduction corresponds to the constraints

\begin{equation}
L =M=N=1,
\end{equation}
\begin{equation}
A+B =0,\quad C+D=0,\quad E+G=0.
\end{equation}
In this case the DCS (50-53) becomes

\begin{equation}
\frac{A_{3}}{A} =\frac{C_{2}}{C}=\frac{E_{1}}{E},
\end{equation}
\begin{equation}
A_{3}C+E_{1}C-E_{1}A-1 =0.
\end{equation}

Equations (72) again lead to the expressions (67) for A,C,E.
Equation (73) takes the form

\begin{equation}
\tau _{1}\tau _{23}-\tau _{2}\tau _{13}+\tau _{3}\tau _{12}-\tau \tau
_{123}=0.
\end{equation}
This equation is nothing but the Hirota-Miwa bilinear discrete
equation for the KP hierarchy of B type (BKP hierarchy) [20]. \
So, the Hirota-Miwa equation (74) together with the formulae (67)
describe discrete deformations of the ''algebra'' (47-49) under
the constraints (70-71). In contrast to the previous case these
deformations are not isoassociative.

Solutions of the Hirota-Miwa equations (68) and (74) are given by
the AKP and BKP $\tau $-functions [19,20]. Thus, any $\tau $-
function of the AKP and BKP hierarchies defines discrete
deformations of the structure constants for the corresponding
algebras.

Finally, we note that the linear equations (9) for all three above
reductions give rise to the linear problems for the systems
(60),(68) and (74). For the last two cases they are

\begin{equation}
\left( T_{j}T_{k}+\frac{\tau _{j}\tau _{k}}{\tau \tau _{jk}}%
(T_{j}-T_{k})-\alpha \right) \mid \Psi \rangle =0, \quad j\neq k,
j,k=0,1,...,N-1.
\end{equation}%
where $\alpha =0$ for the AKP case and $\alpha =1$ for the BKP case, that
coincides with the well-known linear problems [21,20].

\section{Conclusion}

Discrete equations and corresponding deformations considered in the paper
represent only a part of the vast variety. Choosing different algebras to
deform and deformation driving algebras, one can get most of the known
discrete equations within the described scheme. For instance , discrete
deformations of the infinite-dimensional algebra in the Faa' de Bruno basis
for which $C_{jk}^{l}=\delta _{j+k}^{l}+H_{j-l}^{k}+H_{k-l}^{j}$ [11-13] \
are described by the discrete KP hierarchy.

\ The results presented above reveal also a deep connection of the theory of
discrete deformations for associative algebras with the algebraic geometry (
Fay's trisecant formulae, addition formulae for $\tau $ -function) and
discrete geometry ( quadrilateral lattices and all that).

\bigskip

\textbf{References} \bigskip

1. Gerstenhaber M.,1964 On the deformation of rings and algebras, Ann.
Math., \textbf{79}, 59-103 .

2. Gerstenhaber M.,1966 On the deformation of rings and algebras. II, Ann.
Math., \textbf{84}, 1-19 .

3. Witten E., 1990 On the structure of topological phase of two-dimensional
gravity, Nucl. Phys., \textbf{B 340}, 281-332.

4. Dijkgraaf R., Verlinde H. and Verlinde E., 1991 \ Topological strings in $%
d\prec 1$, Nucl. Phys., \textbf{\ B 352,} 59-86 .

5. Dubrovin B., 1992 Integrable systems in topological field theory, Nucl.
Phys., \textbf{B 379}, 627-689 .

6. Dubrovin B., \ 1996 Geometry of 2D topological field theories, Lecture
Notes in Math., \textbf{1620}, 120-348 , Springer, Berlin.

7. Hertling C. and Manin Y.I., 1999 Weak Frobenius manifolds, Int. Math.
Res. Notices, \textbf{6}, 277-286 .

8. Manin Y.I.,2005 F-manifolds with flat structure and Dubrovin's duality,
Adv. Math., \textbf{198}, 5-26 .

9. Manin Y.I., 1999 \textit{Frobenius manifolds, quantum cohomology and
moduli spaces}, AMS, Providence, .

10. Hertling C.,\textit{\ 2002 Frobenius manifolds and moduli spaces for
singularities}, Cambridge Univ. Press, .

11. Konopelchenko B.G. and \ Magri F., \ 2007 Coisotropic deformations of
associative algebras and dispersionless integrable hierarchies, Commun.
Math. Phys., \textbf{274}, 627-658 .

12. Konopelchenko B.G. and Magri F., 2007 Dispersionless integrable
equations as coisotropic deformations: extensions and reductions, Theor.
Math. Phys., \textbf{151}, 803-819 .

13. Konopelchenko B.G. 2008 Quantum deformations of associative algebras and
integrable systems, arxiv:0802.3022.

14. Merkulov S.A. \ 2004 Operads, deformation theory and F-manifolds, in:
Frobenius manifolds, quantum cohomology and singularities, ( Eds. C.
Hertling and M. Marcoli), Aspects of Math., \textbf{E36}213-251.

15. Losev A. and Manin Y.I. 2004 \ Extended modular operads, in: Frobenius
manifolds, quantum cohomology and singularities, ( Eds. C. Hertling and M.
Marcoli), Aspects of Math., \textbf{E36} 181-211.

16. Bogdanov L.V. and Konopelchenko B.G. 1995 Lattice and q-difference
Darboux-Zakharov-Manakov system via $\overline{\partial }$-dressing method,
J. Phys. A :Math. Gen.,\textbf{\ 28}, L173-178.

17. Doliwa A. and Santini P. 1997 Multidimensional quadrilateral lattices
are integrable, Phys. Lett. A \textbf{233}, 365-372.

18. Konopelchenko B.G. and Schief W.K. \ Three-dimensional integrable
lattices in Euclidean spaces:conjugacy and orthogonality, Proc.Royal
Soc.London, Ser.A., \textbf{454 }3075-3104.

19. Hirota R, 1981 Discrete analogue of a generalized Toda equation, J.
Phys. Jpn., \textbf{50}, 3785-3791.

20. Miwa T., 1982 On Hirota's difference equations, Proc.Japan Acad.,
\textbf{58A}, 8-11.

21. Date E., Jimbo M. and Miwa T., 1982, Method for generating discrete
soliton equations I, J. Phys. Jpn., \textbf{51, }4116-4127.

\end{document}